\documentclass[namedreferences]{kluwer} 
\usepackage{graphicx}
\begin{document}
\begin{article}

\begin{opening}
\title{THE BEST OBSERVABLES FROM THE POINT OF VIEW OF A MODEL MAKER}
\subtitle{Best Observables}

\author{Francesca \surname{Matteucci}\email{francesc@sissa.it}}
\institute{
Dipartimento di Astronomia \\
           Universita' di Trieste\\
           Via G. B. Tiepolo 11, 34100 Trieste, Italy}

\runningtitle{Best Observables for Models}
\runningauthor{Matteucci}

\begin{abstract}
We select and discuss the best observables to be used to constrain  
models of galactic chemical evolution. To this purpose, we discuss 
the observables in our 
Galaxy, which is the best studied system, as well as in external galaxies.
\end{abstract}



\end{opening}

\section{Introduction}

The ``best observables'' from the point of view of a chemical evolution 
model maker are those  which allow us to significatively constrain the model 
parameters. In other words, the best observables can be reproduced only by a 
specific combination of the model parameters.
In order to analyze these best observables we should describe first 
the main parameters entering chemical evolution models:

\begin{itemize}
\item the birthrate function, namely the star formation rate (SFR) 
and the initial mass function (IMF),
\item the nucleosynthesis and stellar evolution, namely the yields,
\item the gas flows (infall and outflow),
\item the feed-back between supernovae (SNe) and the interstellar medium (ISM).
\end{itemize}

The still embryonic nature of our knowledge of these quantities forces us
to describe them by means of some free parameters
and to make some simplifying
assumptions.

The free parameters used in models of galactic chemical evolution are:

\begin{itemize}
\item the expression for the SFR (the exponent of the gas density and 
the efficiency of star formation),

\item the expression for the infall or outflow rates,
\item the amount of energy injected from SNe into the ISM which 
effectively heats the gas.
\end{itemize}
The stellar nucleosynthesis and the IMF, although still uncertain, 
can be taken by independent sources. Therefore they cannot be considered 
as completely free parameters. 
As a ``golden rule'' we suggest that, in general, a good model of 
galactic chemical evolution should reproduce a basic 
set of observables and the number of free parameters should be less than 
the number of observables that they can fit.

The main assumptions made in models of galactic chemical evolution are:
\begin{itemize}
\item the instantaneous recycling approximation (I.R.A.) 
(i.e. ignoring the stellar
lifetimes),
\item the homogeneous and instantaneous mixing approximation.
\end{itemize}
The I.R.A. is a poor approximation if one wants to follow the evolution 
of the abundances of chemical elements produced on long timescales such 
as Fe and N
or if one wants to model the evolution of a 
region where the gas fraction is small
($<0.1$).
Therefore, most of chemical evolution models nowadays relax the I.R.A. whereas,
on the other hand, the other assumption is still widely used, and is a good approximation if relatively small regions are considered.

In our discussion we will focuse on the observables in our Galaxy, which is the
best studied system.
Some of these observables turn out to be better than others in the sense that
they allow us to really constrain the models, since they can be reproduced 
only by a specific combination of the various parameters.
In order to identify the best observables from the point of view of a model 
maker in the next section we will describe in more detail the main 
observational constraints and their relation to the 
mechanisms of formation and
evolution of the Galaxy. 

For external galaxies the number of observables is smaller: 
we observe chemical abundances and gas distribution also in external spirals, 
whereas in ellipticals we can only measure integrated properties such as colours 
and spectra.
We will try to envisage here which are the best observables also for these galaxies and what we expect to measure in the future.

\section {The observables in the Milky Way}
Let us take our Galaxy as an 
example.
The Galaxy is a spiral system where four main stellar components can be 
envisaged: the stellar halo, the thick disk, the thin disk and the bulge.
The stars in each of these components have different chemical and 
kinematical characteristics; in particular, the stars in the halo are the most 
metal poor ($<[Fe/H]>  \sim$ -1.5 dex),
they travel on elongated orbits and therefore tend to
possess high radial velocities. 
Their scale height above the Galactic plane is of the order of 3 kpc. 
The stars in the thin disk are rotating in circular orbits 
around the Galactic center, have a scale height
above the plane of the order of 200 pc and average metallicity of 
$<[Fe/H]>$ =-0.2 dex. 
The thin disk contains also gas and dust and the gas is either neutral hydrogen 
(HI) or molecular hydrogen ($H_{2}$). 
The thick disk stars have characteristics intermediate 
between the halo and the thin disk stars, both chemically and kinematically.
The average metallicity of the thick disk stars is $<[Fe/H]> =-0.6$ dex
and their scale height is around 920 pc.
The bulge stars have metallicities in the range -1.5 $\le [Fe/H] \le$ +1.0 dex
and kinematics more like that of the halo stellar population.

The basic observables for our Galaxy are:
\begin{itemize}
\item 
1) the relative number of halo and disk stars in the solar 
neighbourhood (S.N.),
\item 2) the metallicity distribution of stars in the halo,
bulge and local disk,
\item 3) the local present day mass function (PDMF),
\item 4) the absolute solar abundances,
\item 5) the age-metallicity relation(s),
\item 6) the relative abundance ratios as functions of the relative metallicity (relative to the Sun),
\item 7) the distribution of angular momentum per unit mass of the stars of the different Galactic components,
\item 8) the present time infall rate, gas fraction and SN rates (Ia, Ib, II)
in the S.N.,
\item 9) the abundance gradients along the disk,
\item 10) the distribution of gas (HI plus $H_{2}$) and SFR along the disk.
\end{itemize}

1) The ratio between the halo stars and the total number of stars in the S.N. is $\sim 0.03$ (Pagel and Patchett 1975). However, this number is probably underestimated and a more realistic value of this ratio is $\sim 0.1$ (Chiappini et al. 1997). Models which attempt to form the disk out of the gas shed by the halo tend to overestimate this ratio.
Therefore, the indication is that the disk
should have formed mainly out of extragalactic gas.

 2) The G (or F)-dwarf metallicity distribution: 
there are less than $10\%$ of stars with [Fe/H]$<$ -1.0 dex and
the distribution peaks at around -0.2 dex (see figure 1).
This distribution is strictly related to the history of the star formation 
in the local disk, namely to the SFR and the IMF.
The SFR, in turn,  depends crucially on the mechanism of formation 
of the Galactic disk.
A good fit of this distribution shows that 
the local disk formed by slow infall, in particular on a timescale of 6-8 Gyr
(Chiappini et al. 1997; Boissier and Prantzos 1999).
Slow infall is, in fact, the best solution to the G-dwarf problem 
(the fact that the Simple Model of chemical evolution predicts too many 
metal poor disk stars).

\begin{figure}
\centerline{\includegraphics[width=14pc]{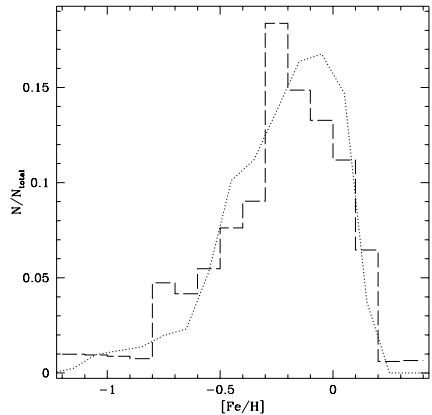}}
\caption{
Observed and predicted G-dwarf metallicity distribution. 
The data are from Rocha-Pinto and Maciel (1996) whereas the model 
(continuous line) is from
Chiappini e al. (1997).}
\label{gdd}
\end{figure}

The metallicity distribution of halo stars 
is different from that of disk
stars and indicates a formation for the halo faster than for 
the disk.
Again, it is not possible to reproduce both distributions with 
a model implying that the disk forms out of the gas lost from the halo.

The metallicity distribution of stars in the Bulge shows a shape more 
similar to that of the halo stars but is skewed towards a much more metal 
rich domain (up to [Fe/H] $\sim$ +1.0 dex).
This distribution (Mc William and Rich 1994) peaks at around 
[Fe/H]=0.0. 
Comparison with theoretical models shows that the Bulge formed faster 
than the disk and with a flatter IMF (Matteucci and Brocato, 1990; 
Matteucci et al. 1999).

\begin{figure}
\centerline{\includegraphics[width=14pc]{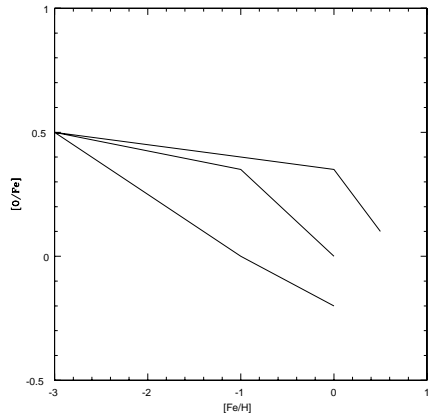}}
\caption{A sketch of the predicted 
behaviour of the [O/Fe] ratio in the framework of the time-delay model for 
three different histories of star formation corresponding to: the upper curve 
with the longest plateau to the bulge, the median curve to the solar 
neighbourhood and the lower curve 
to magellanic irregular galaxies and/or the outermost regions of the Galactic 
disk.} 
\label{sfh}
\end{figure}

3) The local PDMF is the distribution with mass 
of the local  Main Sequence stars. 
It represents an important constraint, 
although very few model makers take it into account, because it represents the 
convolution of the IMF and the SFR. 
A good model of chemical 
evolution should try to reproduce the PDMF since this guarantes 
that there is consistency between the IMF and the SFR.
Most of the uncertainties in the PDMF reside in the low mass end, where data are lacking.

4) The solar and present time
absolute abundances are known with good accuracy. However, the predicted abundances
depend on all the model assumptions and 
therefore they can be reproduced by several combinations of model parameters. 
For 
this reason 
they do not represent an observable which can impose strong constraints 
on models.

5) The age-metallicity relation indicates that [Fe/H] has 
continuosly increased
with the galactic age, although the logarithmic nature of [Fe/H] gives 
the impression of a flattening at late times.
This observable contains many uncertainties due to the uncertain stellar ages
(they can be wrong by a factor of two) 
and it is not a good constraint since it can be reproduced either 
by a closed-box model without I.R.A. or by an infall model with I.R.A.

In addition, a large spread in [Fe/H] is observed at any fixed age
and the nature of this spread is not yet clear.
Therefore, we cannot learn much from the age-metallicity relation.
Finally, one should rather speak of age-metallicity relations since it is
very likely that different Galactic regions had 
different enrichment histories.

6) The abundance ratios  versus metallicity ([el/Fe] vs [Fe/H])
are good observables.
Abundance ratios depend only upon the stellar yields (IMF plus stellar 
nucleosynthesis) whereas the abundance ratios  versus metallicity 
([el/H] vs [Fe/H])
depend upon the yields and the star formation history 
(through [Fe/H]).
Under the assumption of a constant IMF in space and time these
relations can be interpreted in the framework of the time-delay between 
the enrichment due to SN II and SN Ia (time-delay model).
A different star formation history, under this assumption, results in 
a different [el/Fe] versus [Fe/H] relation (see for example [O/Fe] vs. [Fe/H]).
As shown in figure 2, galaxies or Galactic regions with a slower 
evolutionary history show a change in the slope of the [O/Fe] ratio  
occurring at 
smaller metallicities than in  regions where the SFR has been 
quite fast (e.g. the Galactic bulge).
Therefore, the fit of such diagrams gives us an indication about 
the star formation history of the Galaxy, besides the information 
on the nucleosynthesis and SN progenitors.
An interesting aspect of these plots is that the [Fe/O] vs. [O/H] 
and [Fe/Mg] vs. [Mg/H] relations indicate that there has been a period between 
the formation of the halo and the disk when the star formation must 
have stopped. 
This effect is visible from the steep rise of the [Fe/O] at [O/H] 
$\sim$ -0.2 dex,
indicating that for a certain period (models indicate this period 
as no longer than 1 Gyr, Gratton et al. 2000) 
the Fe abundance was increasing whereas the O one was constant
and this can be explained by a halt in the SFR.
The same effect is present in the relation [Fe/Mg] vs. [Mg/H] (see Fuhrmann
1999)

\begin{figure}
\centerline{\includegraphics[width=14pc]{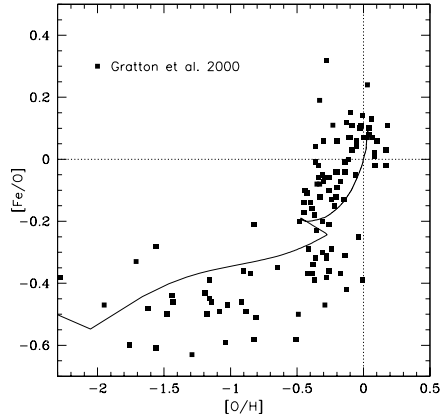}}
\caption{Observed and predicted  [Fe/O] versus [O/H]. The data are from 
Gratton et al. (2000) whereas the models 
are from
Chiappini et al. (1997).}
\label{gap}
\end{figure}

Another interesting finding is by Nissen and Schuster (1997) 
who discovered some halo stars, which are likely to have 
formed in the outer halo,
with metallicities overlapping those of disk stars (-1.3$\le [Fe/H] \le-0.5$) 
but with [$\alpha$/Fe] ratios lower than the corresponding ratios 
in disk stars of the same metallicity. 
As a consequence of this, the transition between halo and disk was 
probably not a smooth one, as originally suggested by Eggen et al. (1962). 
A possible interpretation for these halo stars can be that the halo 
formed inside-out, 
namely that the external halo was formed by slow infall.
An alternative explanation can be that these halo stars have been accreted
from dwarf satellites of the Galaxy, where the star formation proceeded 
in bursts followed by quiescent periods during which only iron was produced.

7) The distribution of the angular momentum per unit mass as a function 
of the angular momentum (Wyse and Gilmore, 1992)  
shows that the halo and the bulge stars have a 
very similar distribution and the same holds for the stars in the thick 
and thin disk but the
distribution of the halo-bulge is different from that of the thick-thin 
disk, clearly indicating that the bulk of the disk cannot have formed out 
of gas lost by the halo.
As a consequence of this, we conclude that it is likely that
most of the gas which formed 
the disk came from outside.

8) The present time infall rate, gas fraction and SN rates 
are good constraints only if
coupled with all the others. In fact, 
theoretical models can always be adjusted to reproduce these absolute
quantities by
varying the various parameters, as in the case of the absolute abundances.
However, the ratio between the SN rates is a good observable since it depends 
only on the supernova progenitors and the IMF.

9) Strong constraints on the mechanism of formation of the Galactic disk
are imposed by the 
abundance gradients. The gradients show that the abundances 
of heavy elements tend to decrease with the galactocentric distance. 
Abundance gradients are strongly related to the mechanism of formation 
of the disk, in particular on the infall law and the star formation rate.
It can be shown that the first condition required to fit the observed 
gradients (measured from HII regions, planetary nebulae, B stars) 
is to assume 
an
``inside-out'' formation for the Galactic disk, in the sense that the 
inner disk must have formed faster than the outer disk.
The second condition is that the SFR should be strongly declining with the 
increasing galactocentric distance. Both these assumptions are necessary.
Radial flows may enhance the predicted gradients but only under 
specific conditions.
A particularly useful abundance gradient to measure is the D gradient, which 
is expected to be positive. This element, in fact, is only destroyed in stars
(astration), so it is a good indicator about the star formation history along
the disk.
Gradients of abundance ratios such as [O/Fe] and [N/O] are important since 
they give us an idea about the timescales of disk formation at different 
galactocentric distances. A negative [O/Fe] gradient, for example, 
would mean that the outer regions of the disk formed more slowly than the 
inner one thus allowing more pollution from type Ia SNe for the same [Fe/H].
On the other hand, a positive [O/Fe] gradient would mean the contrary.

10) The conclusions above are valid also for reproducing the SFR and the 
gas distributions along the disk.
These distributions cannot be reproduced unless a strongly varying SFR is 
assumed. 
This strong variation is achieved either by adding a dependence of the 
SFR on the total surface mass density or a dependence on $R^{-1}$ ($R$ is the 
galactocentric distance (see Prantzos and Boissier, 1999).
The dependence on the surface gas density which best fits the data is 
$k \sim 1.5$, in very good agreement with the observational estimate 
by Kennicutt (1998).
The gas distribution along the disk shows a maximum 
at $\sim 4$ kpc followed by a rapid drop for smaller 
galactocentric distances,
which can be better explained by including some 
dynamical effects such as the presence of a 
central bar (see Portinari and Chiosi 2000).
Observationally, the SFR  along the Galactic disk is a quite 
uncertain quantity derived from the distributions of pulsars, SN 
remnants, Lyman continuum photons and molecular clouds under the 
assumption of an IMF.
Therefore, it is better to plot the ratio $SFR(R)/SFR(R_{\odot})$ in order
to avoid the uncertainty related with the choice of the IMF.
The gas distribution along the disk is determined from the HI and 
$H_2$ gas. The distribution of this latter is quite uncertain since 
is derived by 
assuming a conversion factor (usually
constant along the disk) between the amount of CO and the amount of $H_2$.

\section {The observables in spiral galaxies}

In external spirals the most obvious observables are represented by the 
abundance gradients and the gas and SFR distributions, when available.
Abundance gradients and gas distributions in external spirals are similar 
to those in the Milky Way (Henry and Worthey, 1999). 
This fact may indicate that galactic disks have a common origin
(i.e. an inside-out formation) and that we can perhaps just describe 
the evolution of disks by using suitable
scaling laws based on differences on the total and gas mass.
Other observables in external spirals and in the Galaxy are the colour 
gradients along the disk.
Prantzos and Boissier (2000) have shown that 
the assumption of an ``inside-out'' formation for disks can well 
reproduce the observed gradients including the absence of colour gradients at 
large galactocentric distances.
In fact, this hypothesis implies different scale lengths for the
distribution of stellar profiles.
In particular, in the inside-out scenario the scale lenght 
in the B band is predicted to be
$R_B=4$ kpc
whereas that in the K band is $R_K=2.6$ kpc, in good agreement with 
observations. This is due to the fact that in the inner regions of 
the disk one predicts considerably 
older stellar populations relative to the outer regions where  
there are mostly young stars.
However, photometric models still contain many uncertainties mainly 
because of the existence of the age-metallicity degeneracy problem, 
consisting in the fact that age and metallicity act in the same way on 
integrated colors. Therefore, it is difficult to disentangle the two effects
and integrated colors are not themselves good observables 
unless other constraints are considered at the same time.

\section{Observables in elliptical galaxies}

The observables which are relevant to study the chemical evolution of 
elliptical galaxies are represented by the metallicity indices, such as 
for example $Mg_{2}$,  $<Fe>$ and $H_{\beta}$. These indices are measured 
from integrated spectra and depend on the metallicity and the age 
(again the age-metallicity degeneracy)
of the stellar population which dominates in the visual light.
The index $H_{\beta}$ perhaps is an exception since it depends mostly on the age and it can be used to break the degeneracy.
The main problem with metallicity indicators is that they do not represent the real abundances and one needs to use a calibration to 
calculate such abundances.
Elliptical galaxies show a mild increase of $Mg_{2}$ with  
galactic mass (measured through the stellar velocity dispersion): this 
is known as ``mass-metallicity'' relation.

Data for cluster and field ellipticals have shown that the relation 
$<Fe>$ vs.
$Mg_{2}$ is quite flat, indicating that the [Mg/Fe] ratio should 
be an increasing function of the galactic luminosity.
This finding is the contrary of what is expected from galactic models 
with SN-driven galactic winds, as originally proposed by Larson (1976).
These models, in fact, predict exactly the opposite, 
due to the longer star formation period predicted for the more massive 
ellipticals relative to the less massive ones.
An explanation for this behaviour can be that galactic winds are occurring 
earlier in massivegalaxies than in small galaxies 
and this can be achieved if the SFR 
increases with galactic mass. However, this trend of the SFR has to 
be calibrated in such a way not to destroy the mass-metallicity relation
(see Matteucci 1994).
The metallicity indices indicate the existence of abundance gradients also inside ellipticals but it is not yet clear if the gradient of $Mg_{2}$ is flatter,
steeper or the same as the gradient of $<Fe>$.
This is an important point in order to understand the formation 
and evolution of these galaxies.
Another strong constraint on the evolution and formation 
of ellipticals is provided by the abundances and 
abundance ratios measured in clusters of galaxies 
(see Matteucci 1996 for a review).

\section{Conclusions}

In this paper we have analyzed the best observables from the point 
of view of a model maker. 
Although we are not able to identify a unique model of galactic chemical 
evolution, we can envisage those observables which allow us to better 
constrain 
the models.
We have discussed the available 
constraints for the Milky Way, external spirals and ellipticals and
their interpretation 
by means of chemical evolution models.
The comparison between models and observations suggests 
that the best observables are:
\begin{itemize}

\item i) The abundance ratios and the relation between abundance ratios 
and metallicity. They allow us to impose constraints on the stellar 
nucleosynthesis and on the star formation history, under the 
assumption of a constant IMF.

\item ii) The distribution of dwarf stars as a function of metallicity.
This 
is related to the star formation history and therefore to the evolution of 
the gas content. It would be auspicious to be able to observe such 
a distribution
also in external galaxies, both spirals and ellipticals to gain 
insight onto their formation and evolution.

\item iii) Kinematical and chemical studies of stars representing the 
halo-disk transition  provide a very important constraint in connection
with 
the formation and evolution of the Galaxy. 
In particular, it is important to search for possible correlations between
orbital parameters and abundance ratios.

\item iv) The abundance gradients along galactic disks are sensitive to 
the star formation history and stellar nucleosynthesis. 
They can be used to infer the story of the formation of disks and
gradients of abundance ratios between elements, formed on different timescales, 
can also give an idea about the
timescales of disk formation at various galactocentric distances.
Abundance gradients are now measured in external spirals and in 
elliptical galaxies (although in this case the metallicity is inferred 
through indices) and we hope that in the next years more and more data will be 
available.
In particular, we would like to see detailed abundance 
measurements of elements such as  $\alpha$-elements 
(O, Mg, Si, Ca) and Fe  which represent an important tool for understanding 
the 
mechanisms of galaxy formation and evolution in external galaxies, since 
they can be used as 
cosmic clocks and are related to the star formation history.
As a consequence, abundances and abundance ratios can also be used to infer 
the nature of high redshift galaxies.

\end{itemize}

\begin{acknowledgements}
I would like to thanks Cristina Chiappini and Donatella Romano
for their contribution to  the work describe here. 
This work has been partially supported by the Italian MURST, through
COFIN98 at Padova.
\end{acknowledgements}

\end{article}
\end{document}